\documentclass[twocolumn,showpacs,preprintnumbers,amsmath,amssymb,letter]{revtex4}

\usepackage{graphicx}
\usepackage{dcolumn}
\usepackage{bm}

\begin{document}

\title{New $T=1$ effective interactions for the 
f$_{5/2}$ p$_{3/2}$ p$_{1/2} $ g$_{9/2}$ model space; \\
 Implications for valence-mirror symmetry and seniority isomers}
        
\author{A.~F.~Lisetskiy$^{1}$, B.~A.~Brown$^1$, M.~Horoi$^{2}$, and H.~Grawe$^3$}

\affiliation{$^1$ National Superconducting Cyclotron Laboratory, Michigan State University,
                 East Lansing, Michigan 48824-1321}

\affiliation{$^2$ Physics Department, Central Michigan University, Mount Pleasant, 
Michigan 48859}

\affiliation{$^3$ Gesellschaft f\"ur Schwerionenforschung mbH, D-64291 Darmstadt,
Germany}

\date{\today}

\begin{abstract}
New shell model Hamiltonians  are 
  derived for the $T=1$ part of the 
residual interaction in the f$_{5/2}$ p$_{3/2}$ p$_{1/2}$ g$_{9/2}$ model space 
based on the 
analysis and fit of the available experimental data for 
$^{57}_{28}$Ni$_{29}$--$^{78}_{28}$Ni$_{50}$ isotopes and $^{77}_{29}$Cu$_{50}$--
$^{100}_{50}$Sn$_{50}$ isotones. The fit procedure, properties of the determined 
effective interaction as well as new results for valence-mirror symmetry 
and seniority isomers for nuclei near $^{78}$Ni 
and $^{100}$Sn are discussed.  
\end{abstract}


\maketitle          


Neutron-rich nickel isotopes in the vicinity of $^{78}_{28}$Ni$_{50}$ are currently 
in the focus of modern nuclear physics and astrophysics studies 
\cite{Bro95,Grz98,Graw02,Saw03,Ish00,Sor02,Lang03}.  
The enormous interest in
this region is motivated by several factors.  The primary issue concerns the doubly magic 
nature of  $^{78}_{28}$Ni  and understanding the way in which the 
neutron excess will affect the properties of nearby nuclei and the  $^{78}$Ni core itself.
The shell-model orbitals for neutrons in nuclei with $Z=28$ and N=28-50 
($^{56}$Ni-$^{78}$Ni) are the same as those for protons in nuclei with N=50 and Z=28-50
($^{78}$Ni-$^{100}$Sn). Thus it is of interest to understand 
the similarities and differences in the properties of these nuclei with 
valence-mirror symmetry (VMS)  \cite{Wir88}.
The astrophysical importance is related to the understanding of the nuclear mechanism 
of the rapid capture of neutrons by seed nuclei through the r-process. The path of this reaction network 
is expected in neutron-rich nuclei for which there is little experimental data,
and the precise trajectory is dictated by the 
details of the shell structure far from stability. 

Experimental investigations of neutron-rich nuclei have greatly advanced the last 
decade providing access to many new regions of the nuclear chart.
Nuclear structure theory in the framework of shell-model configuration mixing has also
advanced from, for example, the elucidation of the properties of the sd-shell nuclei 
($A=16-40$) in the 1980's \cite{sd} to those of the pf-shell ($A=40-60$)  in current 
investigations \cite{pf1,pf2,pf3}. Full configuration mixing in the next oscillator shell 
(sdg) is presently at the edge of computational feasibility. 
For heavy nuclei the spin-orbit interaction pushes the $g_{9/2}$ and $f_{7/2}$ orbits 
down relative to the lower-$l$ orbits.
Thus the most important orbitals for neutrons in the region of
$^{68}$Ni to $^{78}$Ni are $p_{3/2}$, $ f_{5/2}$, $p_{1/2}$ and $ g_{9/2}$ 
(referred to from now on as the $pf_{5/2}g_{9/2}$ model space). It is noteworthy that 
this model space is not affected by center-of-mass spurious components.  
Full configuration mixing 
calculations for neutrons or protons in this model space are relatively easy. 
The work we describe here on the $T=1$ effective interactions will provide a part 
of the input for the larger model space of both 
protons and neutrons in these orbits where the maximum m-scheme dimension is 
13,143,642,988. This proton-neutron model space is computationally feasible with 
conventional matrix-diagonalization techniques for many nuclei in the mass region A=56-100, 
and Quantum Monte Carlo Diagonalization techniques \cite{Monte} or Exponential 
Convergence Methods \cite{Hor03} can be used  for all nuclei. 

The present paper reports on new effective interactions for the
$pf_{5/2}g_{9/2}$ model space derived from a fit
to experimental data for Ni isotopes from $A=57$ to $A=78$ and $N=50$ isotones 
from $^{79}$Cu to $^{100}$Sn for neutrons and protons, respectively. Predictions 
for the $^{72-76}$Ni isotopes are made using the new effective interaction.
For the first time the calculated structures of the $^{68,70,72,74,76}$Ni isotopes  and the 
$^{90}$Zr,  $^{92}$Mo,  $^{94}$Ru,  $^{96}$Pd, $^{98}$Cd are compared and analyzed 
with respect to the VMS concept \cite{Wir88}.
Our work provides a much improved Hamiltonian for $Z=28$ over those considered in 
smaller model spaces \cite{Bro95,Grz98,Graw02,Graw97},
and also provides a new Hamiltonian for $N=50$ that is similar to those obtained  
previously \cite{Ji88_effint,Sin92}.

The effective interaction is specified uniquely in terms of interaction parameters 
consisting of four single-particle energies
and 65 $T=1$ two-body matrix elements (TBME).
The starting point for the fitting 
procedure was a realistic G-matrix interaction based on the Bonn-C $NN$ potential
together with core-polarization corrections based 
on a $^{56}$Ni core \cite{HJen_private}.  The low-energy levels known 
experimentally are not  
sensitive to all of these parameters, and thus not all of them can be well determined by the selected 
set of the energy levels. Instead, they are sensitive to certain linear combinations of the 
parameters. The weights and the number 
of the most important combinations can be found with the Linear 
Combination Method (LCD) \cite{Honma_LCD}.  Applying LCD for our fit we found that 
convergence of the $\chi^2$ in the first iteration is achieved already at 20 linear combinations 
and we have chosen this as a reasonable number for all following iterations. 
We performed iterations (about six) until the eigen-energies converged.
\begin{figure}
\includegraphics[scale=0.34]{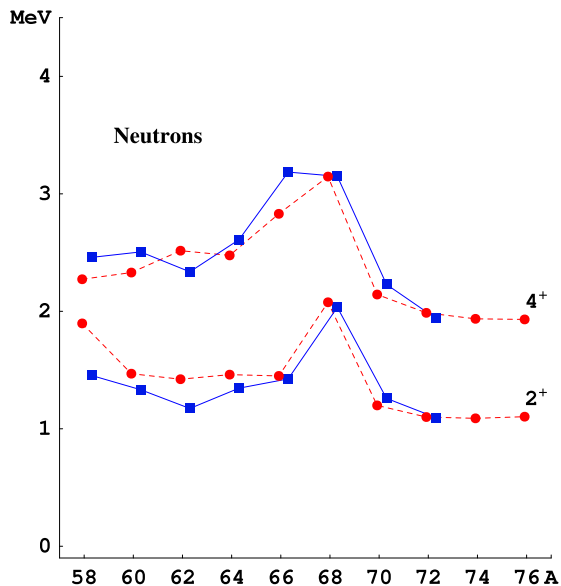}
\vspace{-1.3 cm}
\caption{Calculated and experimental excitation energies of the $2^+_1$ and $4^+_1$ states 
in $A=58-76$ even-even nickel isotopes.   Calculated levels are
given by circles connected by the dashed line. Experimental data are depicted by 
squares connected by the solid line. }
\label{2and4n}
\end{figure} 
The values of the neutron interaction parameters are 
adjusted to 
fit 15 experimental  binding energies for $^{57-78}$Ni and 91 energy levels for 
$^{60-72}$Ni. The nuclei below $^{60}$Ni were not emphasized in the fit due to the
increased role of excitations from the $f_{7/2}$ orbit as $^{56}$Ni is approached \cite{pf3}. 
In the absence of experimental data on the binding 
energy of nuclei near $^{78}$Ni we include the SKX \cite{skx} Hartree-Fock value of 
-542.32 MeV for the binding energy of $^{78}$Ni as a "data" for the fit. Our calculated 
binding energies for $^{73-77}$Ni isotopes agree well with the recent corresponding
extrapolations from Ref. \cite{Aud03}. 
For protons 19 
binding energies and 113 energy levels were used in a fit (this data set is 
similar to that used in Ref. 
\cite{Ji88_effint}).  The average deviation in binding and 
excitation energies between experiment and theory is 241 keV and 
124 keV for neutrons and protons, respectively.  A detailed report of the 
new interactions will be given elsewhere \cite{Lis04_int}.

In this letter we emphasize some of the interesting results for 
known nuclei and the extrapolation to properties of unknown nuclei.
To illustrate some general properties of the new interactions we  plot the excitation 
energies of  the $2^+_1$ and the $4^+_1$ states for neutrons and protons in 
Figs. 1 and 2,  respectively.   The systematics shows good agreement 
between shell-model calculation and experiment. There is some similarity in the trends 
for the nuclei with $A=68-76$ and $A=90-98$, that is referred to  as the 
VMS \cite{Wir88}.  However, the left part of Figs. 1 and 2  
 ($A=58-66$ for nickel isotopes and $A=80-88$ for $N=50$ isotones) are drastically 
different.  Two nuclei, $^{66}$Ni and  $^{88}$Sr, show the most profound 
differences  in the location of the $4^+_1$ state.  The energy gaps between 
the  $4^+_1$ and the $2^+_1$ states in  the $^{62}$Ni and the $^{84}$Se are also 
 obviously distinct.

  \begin{figure}
\includegraphics[scale=0.34]{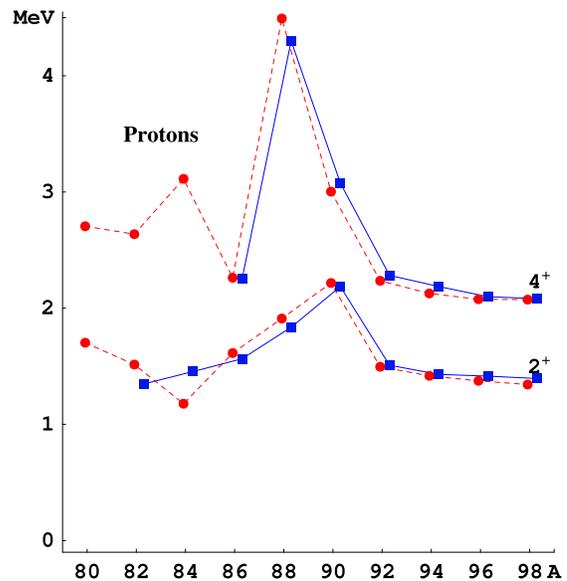}
\vspace{-1.3 cm}
\caption{The same as in the Fig. \protect\ref{2and4n} for 
$A=80-98$ even-even $N=50$ isotones.
}
\label{2and4p}
\end{figure} 
\begin{figure}
\includegraphics[scale=0.31]{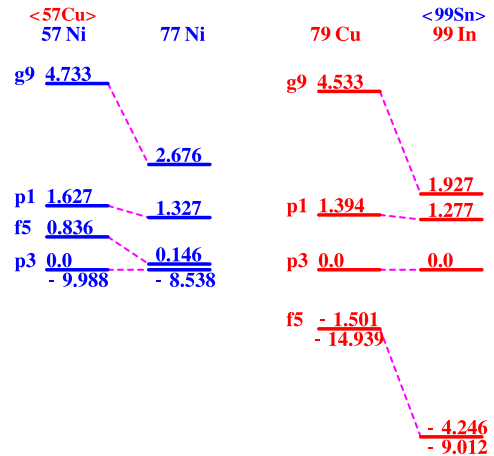}
\vspace{-1.5cm}
\caption{The neutron ($^{57}$Ni)  and proton  ($^{79}$Cu) single-particle
energies (SPE) 
relative to the $^{56}$Ni  and  the $^{78}$Ni cores, respectively.  
The SPE for neutron holes in $^{78}$Ni and proton holes in $^{99}$In
are also shown. The SPE values relative to the 
corresponding cores ($p_{3/2}$ orbital) are given below (above) the plotted lines.
The relative SPE for $^{57}$Cu and $^{99}$Sn are similar
to those of the mirror nuclei $^{57}$Ni and $^{99}$In, respectively.
}
\label{spe}
\end{figure} 

These differences may be qualitatively understood
from the ordering of the single-particle energies (SPE) for both cases (see Fig. \ref{spe}). 
For neutrons  the lowest orbital is $p_{3/2}$, which is followed by the 
 $f_{5/2}$, $p_{1/2}$ and $g_{9/2}$ orbitals. This ordering is similar to the familiar 
cases of interactions in the $pf$-shell. For the protons we obtain the   
$f_{5/2}$ orbital as the lowest  similar to the previous Ji and Wildenthal 
interaction \cite{Ji88_effint}. One notes that the spacing 
between $p_{3/2}$, $p_{1/2}$ and $g_{9/2}$ is rather similar in both cases.  
The fact that the $f_{5/2}$ orbital is pushed down in energy in $^{79}$Cu
relative to $^{57}$Cu 
may be attributed 
to the neutron mean field of the $^{78}$Ni core.  The strongly binding monopole 
interaction in the proton-neutron ($\pi \nu$) spin-flip configuration $\pi f_{5/2} \nu g_{9/2}$
as compared to  $\pi p_{3/2} \nu g_{9/2}$ causes a dramatic down sloping of the $\pi f_{5/2}$ 
level in $Z=29$ (Cu) isotopes upon filling of the $\nu g_{9/2}$  orbit \cite{20a,20b}.

The difference in the ordering of the proton 
orbitals as compared to neutrons is the main reason for the 
differences (e.g. for the $4^+$ states) observed in Figs. 1 and 2.
The SPE's impact ground states as well: the $f_{5/2}^6p_{3/2}^4$ component 
($f_{5/2}$ and $p_{3/2}$ are filled) constitutes 59.8$\%$ for  the $^{88}$Sr and only 
21.4$\%$ for the $^{66}$Ni (the VMS partner of  the $^{88}$Sr). This difference determines 
what happens beyond the $^{66}$Ni ( e.g. $^{68}$Ni, 
see also Refs. \cite{Sor02,Lang03}) or  $^{88}$Sr upon filling  the 
$p_{1/2}$ and the $g_{9/2}$ orbitals.

\begin{figure}
\includegraphics[scale=0.38]{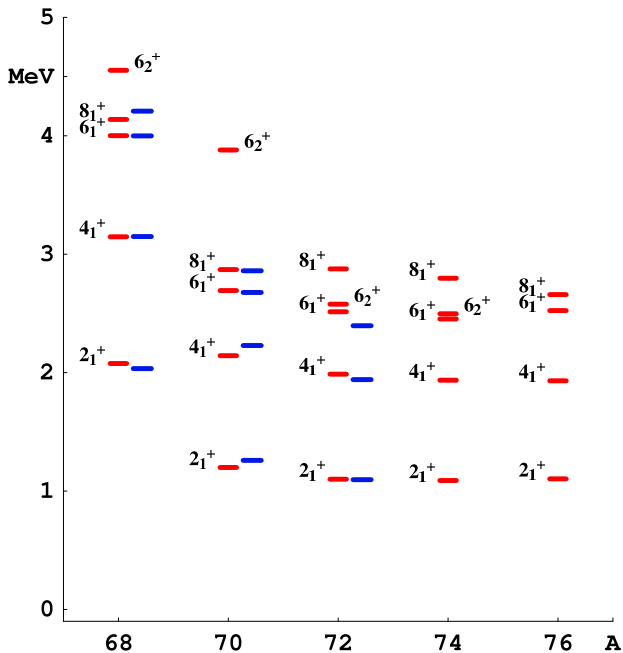}
\vspace{-1.4cm}
\caption{Yrast level schemes for even-even Ni isotopes with $A=68-76$. 
Calculated levels on left side and experimental levels on the right side. }
\label{fig3}
\end{figure}

\begin{figure}
 \includegraphics[scale=0.38]{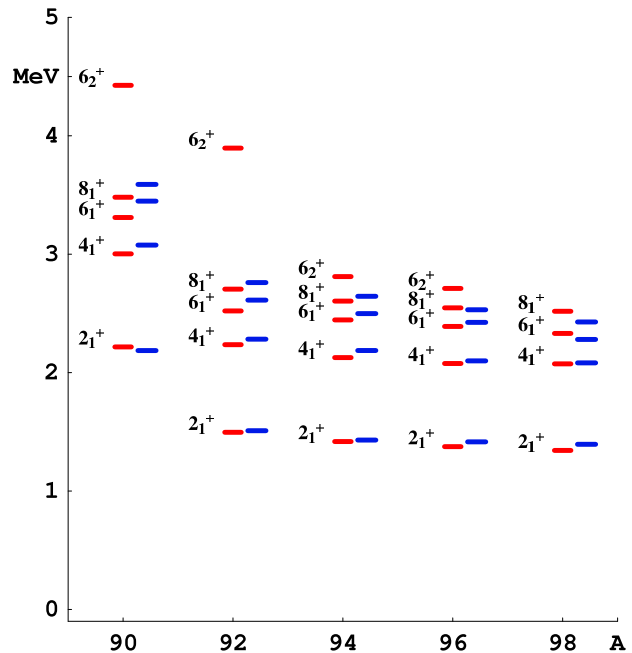}
\vspace{-1.4cm}
\caption{ The same as in the Fig. \protect\ref{fig3} for
even-even $N=50$ isotones with $A=90-98$. 
 }
\label{fig4}
\end{figure}

To compare the details of the low energy spectra of the 
even $^{68-76}$Ni  isotopes and even $A=90-98$  $N=50$ isotones we show the 
calculated
and experimental energies for some levels of interest in Figs. \ref{fig3} and \ref{fig4}, 
respectively.   
The energies of the $2^+_1$,  $4^+_1$,  $6^+_1$ and $8^+_1$
states are, approximately,  the same in all four $^{70-76}$Ni nuclei.
A similar situation holds for the four isotones $^{92}$Mo-$^{98}$Cd, where the validity 
of the generalized seniority approximation is well established \cite{Ji88_trans}.
Indeed, the calculated structure of the wave functions  indicate 
a large contribution of the $g_{9/2}$ orbital  for these nuclei. 
However, this contribution is not so large to conclude the dominance for all 
four nuclei. This is especially well illustrated by the structure of the ground states:
the $[g_{9/2}]^{A-68}_{0^+}$ component in the $0^+_1$ wave functions for 
$^{70,72,74,76}$Ni   is  44 $\%$, 53$\%$, 67$\%$ and 83$\%$, respectively. It is obvious 
that, in contrast to the single $g_{9/2}$ orbital approximation,  the structures of $^{70}$Ni  
and $^{76}$Ni are significantly different. The contributions of the $[g_{9/2}]^{A-68}_{2^+}$ 
component to the $2^+_1$ states have  approximately the same weight.  The other 
components of the wave functions play a very important role. For instance, the 
difference between the effective neutron $g^2_{9/2};J=0$ and  $g^2_{9/2};J=2$  
TBME's  is 0.373 MeV, however the $2^+-0^+$ energy gap in  $^{70}$Ni or in $^{76}$Ni
is 1.2-1.1 MeV.  Thus the largest contribution to the gap (0.8-0.7 MeV)  is mixing
with other configurations. The 
$g_{9/2}$ wave function content of the corresponding valence mirror partners 
 $^{92}$Mo-$^{98}$Cd is slightly larger :  51 $\%$, 60$\%$, 71$\%$ and 84$\%$ of 
$[g_{9/2}]^{A-90}_{0^+}$ in the ground states of each A-isotone, respectively, but 
the overall situation is rather similar to $^{70,72,74,76}$Ni. 

The energies of the 
$2^+_1$ states for the nickel isotopes are systematically lower ($\sim 0.3$ MeV) than for 
the $N=50$ isotones. This effect is due to the properties of effective 
$g^2_{9/2}$ TBME's,  since the energy gap between  the effective $g^2_{9/2};J=0$ and  
$g^2_{9/2};J=2$  TBME's  is  0.727 MeV and 0.373 MeV for protons and neutrons,
 respectively ( i.e. they differ by approximately the same $\sim 0.3$ MeV) and 
the calculated structure of the $^{70-76}$Ni and $^{92}$Mo-$^{98}$Cd is similar
(the energy gap in the starting renormalized Bonn-C Hamiltonian is 0.538 MeV).
We interpret this as an indication that the $Z=28$ proton shell
gap near $^{68}$Ni is relatively weak compared to the $N=50$ neutron shell gap
near $^{88}$Sr. Thus, the nuclei $^{56-78}$Ni have substantial amounts of proton
core excitations that are not included explicitly in the model space, but
are implicitly taken into account by the effective TBME. One would like to treat 
the nickel isotopes in a model space which explicitly includes
the proton excitations, but this is presently not feasible in terms of 
computational power.

The lowering of the effective $J=2$-$J=0$ gap for the nickel isotopes 
compared to $N=50$ has important consequences for non-yrast states. This is 
illustrated by the 
properties of the $6^+$ states.  For $^{94}$Ru and $^{96}$Pd the second $6^+$ state 
is dominantly seniority $\nu=4$ and lies above the $8^+_1$ state, while in 
$^{72}$Ni and $^{74}$Ni it is well below the $8^+_1$ state and is
almost degenerate with the 
$6^+_1$ seniority $\nu=2$ state.  Despite the very small splitting between two $6^+$ states
with dominant seniority $\nu=2$ and $\nu=4$  they are only slightly mixed.
The structure and location of the $6^+$ states has important implications for the 
isomeric properties of the $8^+$ states. The 
B(E2; $8^+_1 \rightarrow 6^+_{1,2}$) values are given in Table \ref{be2n}. The 
E2 transition between  the $8^+$ and $6^+$ states of the same seniority is 
forbidden in the  middle of the shell.
\begin{table}
\caption{Calculated B(E2; $J^\pi_i \rightarrow J^\pi_f)$ values for the $A=70-76$ Ni isotopes. 
A reasonable value of 1.0 is assigned, tentatively, to an effective quadrupole charge $e_n$ . 
B(E2) values are given in units of e$^2$fm$^4$. 
Calculated and available experimental lifetimes $\tau$ for the $8^+_1$ state are given 
in last two rows.
 Experimental excitation energies were used in lifetime calculations for $^{70}$Ni . 
For $^{72,74,76}$Ni isotopes theoretical excitation energies were used. }
\label{be2n}
\begin{center}
\begin{tabular}{c|c|cccc}
\hline 
 $J_i^\pi$ & $J_f^\pi$ & $^{70}$Ni  & $^{72}$Ni & $^{74}$Ni  & $^{76}$Ni  \\ \hline
 $2^+_1$  & $0^+_1$ &   64  &  84 &   76 &   46    \\
$4^+_1$  & $2^+_1$ &    51 &   94 &   85 &   54    \\
 $6^+_1$  & $4^+_1$ &   31 &    29  &   34 &     37   \\
 $8^+_1$  & $6^+_1$ &   12 &    1.9 &      9.2 &    15     \\
                & $6^+_2$ &    3.3  &    52 &   47 &   -   \\
\hline
$\tau(8^+_1)$ & Th. & 326.0 ns & 6.1 ns     & 5.1 ns &1.2 $\mu$s  \\
                 &  Expt. & 335(4)\footnote{\vspace{-0.5cm} Ref. \protect\cite{Grz98}; $^b$Ref.  \protect\cite{Saw03};}  ns & $<$26$^b$ ns  & $<$87$^b$ ns  & 
             \\
\hline
\end{tabular}
\end{center}
\vspace{-0.2cm}
\end{table}
It is well known that this seniority selection rule leads to the 
isomerism of $8^+_1$ states in   $^{94}$Ru and $^{96}$Pd \cite{nndc}, with  
measured lifetimes $\tau$ of 102(6) $\mu$s and 3.2(4) $\mu$s, respectively. 
Our calculations result in  lifetimes of the order of $\mu$s as well. The discrepancy between 
theoretical and experimental B(E2) values for the  $^{94}$Ru is relatively large, 
and this is common to previous calculations \cite{Sin92,Ji88_trans}.
\begin{table}
\caption{Calculated and experimental B(E2;$J^\pi_i \rightarrow J^\pi_i-2)$ values for 
$A=92-98$, $N=50$ isotones. A reasonable value of 2.0 is assigned, tentatively, 
to an effective  quadrupole charge $e_p$ \protect\cite{Ji88_trans}.  
B(E2) values are given in e$^2$fm$^4$ units. }
\label{be2p}
\begin{center}
\begin{tabular}{c|cc|cc|cc|cc}
\hline 
 $J_i^\pi$ & \multicolumn{2}{c}{$^{92}$Mo} & 
 \multicolumn{2}{c}{$^{94}$Ru} &   \multicolumn{2}{c}{$^{96}$Pd}  &  \multicolumn{2}{c}{$^{98}$Cd}  \\ \hline
     &   Th. & Expt.\footnote{\vspace{-0.7cm} Ref. \protect\cite{nndc}; $^b$ Ref. \protect\cite{new23}} 
& Th. & Expt.$^a$ & Th. & Expt.$^a$ & Th. & Expt.$^b$ \\ \hline 
 $2^+_1$  &    235  &    207(12) &  304  & -         & 283  & -          &  181 & -\\
$4^+_1$   &   164   &   $<$605  &      9.2  &  -        &  40   & -         &  214 & - \\
 $6^+_1$  &   110   &       81(2)  &     8.2  & 2.9(1)   &  20   & 20(3)   &  149 & - \\
 $8^+_1$  &     42   &    32.4(5)  &     2.7  & 0.09(1) &  7.1     & 8.9(12) & 60    &
35(11)  \\
\hline
\end{tabular}
\end{center}
\vspace{-0.2cm}
\end{table} 
  
Turning back to Ni-isotopes, one notes, that pushing 
down of the $6^+_2$ $\nu=4$ states opens up a  new channel for the fast E2 decay of the 
$8^+$ states that results in the disappearance of isomeric states  in 
$^{72,74}$Ni --the corresponding lifetimes  move down to the ns region (see Table \ref{be2n}).
  Our results with the newly derived interactions fully support the above 
 explanation of the absence of an isomeric  $8^+$ state in $^{72}$Ni proposed in 
\cite{Graw02,Saw03} on the basis of single $g_{9/2}$ orbital shell-model calculations since our 
calculated  wave functions of   the $8^+,6^+$ states in $^{72,74}$Ni are dominated by 
the $g^4_{9/2}$ (70$\%$) and  $g^6_{9/2}$ (85$\%$) configurations, respectively. However, 
it is shown above that this is not the case for the $0^+_1$ and $2^+_1$ states. 

Clearly more detailed experimental studies of the nuclei in the vicinity of the  $^{78}$Ni are required 
to verify the shell-model predictions.  The next step  will be to
combine these new neutron-neutron and proton-proton effective Hamiltonians with a proton-neutron Hamiltonian to describe the wide variety of spherical, 
collective and co-existing features for the $A=56-100$ mass region as well as to apply
the wavefunctions to the calculations of weak-interaction and astrophysical 
phenomena. 


We acknowledge support from NSF grant PHY-0244453.

\end{document}